\documentclass[aps,prl,twocolumn,noshowpacs,10pt]{revtex4-1}

\usepackage{amsmath}
\usepackage{amsfonts}
\usepackage{amssymb}
\usepackage{hyperref}

\begin{document}
\title{Fractional Mirror Symmetry}
\author{Dan \surname{Isra\"el}}
\email{israel@lpthe.jussieu.fr}
\affiliation{Sorbonne Universit\'es, UPMC Univ Paris 06, UMR 7589, LPTHE, F-75005, 
Paris, France}
\affiliation{CNRS, UMR 7589, LPTHE, F-75005, Paris, France}
\begin{abstract}
Mirror symmetry relates type IIB string theory on a Calabi--Yau 3-fold to type IIA on the mirror CY manifold, whose 
complex structure and K\"ahler moduli spaces are exchanged.  
We show that the mirror map is a particular case of a more general quantum 
equivalence, {\it fractional mirror symmetry}, between Calabi--Yau compactifications and non-CY ones. As was done by 
Greene and Plesser for mirror symmetry, we obtain these new  dualities by considering orbifolds of Gepner models, 
of asymmetric nature, leading to superconformal field theories isomorphic to the original ones, but with a
 different target-space interpretation. The associated Landau--Ginzburg models involve both chiral and twisted chiral 
multiplets hence cannot be lifted to ordinary gauged linear sigma-models. 
\end{abstract}
\maketitle

\section{Introduction}

Mirror symmetry plays a major role in our understanding of Calabi--Yau (CY) manifolds, both in their physical and mathematical 
aspects~\cite{Yau:1998yt,Greene:1997ty}. It generalizes  T-duality to CY sigma-models 
as one exchanges the axial and vector R-symmetries of the  $\mathcal{N}=(2,2)$ superconformal 
algebra~\cite{Lerche:1989uy}. The first concrete realization was obtained by Greene and Plesser~\cite{Greene:1990ud} using Gepner 
models~\cite{Gepner:1987qi}; they have shown that an orbifold by the largest subgroup of discrete symmetries preserving 
spacetime supersymmetry (bar permutations) gives an isomorphic conformal field theory (CFT) 
with reversed right-moving R-charges. It provides an equivalence between type 
IIA compactified on some CY and type IIB  on a topologically distinct one, 
whose Hodge diamonds are 'mirror' to each other.  Using the gauged linear sigma-model (GLSM) description~\cite{Witten:1993yc}, 
that includes the Gepner points in a certain regime, Hori and Vafa gave a proof of mirror symmetry~\cite{Hori:2000kt}.

One can wonder whether there exist generalizations of mirror symmetry~\footnote{In the 
context of heterotic strings, mirror symmetry can be extended to $(0,2)$ models~\cite{Blumenhagen:1996vu}.}, 
in particular dualities that would take the sigma-models out of the realm of CY compactifications.  The inspiration for this  
article originates from a recent work~\cite{Israel:2013wwa} where we described fibrations of $K3$ Gepner models over 
a two-torus in type II, breaking space-time supersymmetry from the right-movers only.  A {\it geometric} compactification of this sort would 
exist if the two connections $\nabla \left(\omega \pm H/2\right)$ appearing in the supersymmetry variations gave  different G-structures, 
requiring non-zero three-form flux; for compact models this is forbidden by the tadpole condition $\int e^{-2\Phi} H\wedge \star H=0$. 
The symmetries of the  $K3$ fiber appearing in the monodromies were actually neither geometric nor mirror 
symmetry. These new non-geometric symmetries of CY quantum sigma-models are the focus of the present work. 

We construct a  class of asymmetric modular invariants for Gepner models in type II, using the simple currents 
formalism~\cite{Schellekens:1990xy}, preserving all space-time supersymmetry from the left-movers,  
while the other half is generically broken. This is made possible by a specific choice of 
discrete torsion, which changes in particular the orbifold action on the K\"ahler moduli. A subclass of these asymmetric 
orbifolds gives models isomorphic as conformal field theories to the original ones, such that the axial and vector R-symmetries 
for a {\it single} minimal model are exchanged. They correspond to 'hybrid' Landau--Ginzburg (LG) orbifolds with both chiral and 
twisted chiral superfields, hence cannot be lifted to Calabi--Yau GLSMs. This is another sign of the non-CY 
nature of these new dual models; as this map can be applied stepwise to each and every minimal model we 
give it the name of {\it fractional mirror symmetry}.  

\section{Simple currents and Gepner models}
We first review the simple current formalism~\cite{Schellekens:1989am,Kreuzer:1993tf} and its relation with Gepner models.  
In a CFT a {\it simple current} $J$ is a primary of the chiral algebra whose fusion with a generic primary 
gives a single primary: $J \star \phi_{\mu} = \phi_{\nu}$. This action defines 
the {\it monodromy charge} of the primary w.r.t. the current, $Q_\imath (\mu)= \Delta(\phi_\mu)+\Delta({J_\imath})
-\Delta (J_\imath \star \phi_\mu)\! \mod 1$;  
two-currents are  mutually local if $Q_{\imath} (J_\jmath)=0$. A modular-invariant partition function 
associated with a simple current extension is:
\begin{equation}
\label{eq:curpart}
Z = \sum\limits_{\mu} \prod\limits_{\imath=1}^{M} \sum\limits_{b^\imath \in \mathbb{Z}_{n_\imath}}
\chi_{\mu} (q)\, \chi_{\mu+ \beta_\jmath b^\jmath} (\bar q)  \
\delta^{(1)} \left( Q_{\imath} (\mu) + X_{\imath \jmath} b^{\jmath} \right)\, , 
\end{equation}
with $J_{\imath} \star \phi_{\mu} = \phi_{\mu+\beta_\imath}$ and $n_\imath$  the length of $J_\imath$.
The symmetric part of the  matrix $X$ is determined by the relative monodromies as 
$X_{\imath \jmath} + X_{\jmath \imath}= Q_{\imath} (J_\jmath)\! \mod 1$, while the antisymmetric part, 
{\it discrete torsion}, should be such that:
\begin{equation}
\label{eq:torsionconst}
\text{gcd} (n_\imath,n_\jmath)\, X_{\imath  \jmath} \in \mathbb{Z}\, .
\end{equation}

A Gepner model for a CY 3-fold is obtained from a tensor product of $r$ $\mathcal{N}=2$ minimal models 
whose central charges satisfy \mbox{$\sum_{n=1}^r c_n = \sum_{n=1}^r (3-6/k_n)= 9$}. 
One needs to project the theory onto states with odd integer left and right R-charges; this 
can be rephrased in the simple currents formalism.  The simple currents of the minimal models are 
primaries with quantum numbers ({\it j}=0,{\it m},{\it s}). They can be grouped together with the 
current for a free fermion into $J=(\sigma_0|m_1,\ldots,m_r|s_1,\ldots,s_r)$ 
where $\sigma_{0}$ is the fermionic $\mathbb{Z}_4$ charge. The Gepner modular invariant is a 
simple current extension with $J_0=(1|1,\ldots,1|1,\ldots,1)$, 
ensuring the projection onto odd-integer R-charges, while 
$\{ J_n = (2|0,\ldots,0|0,\ldots,0,2,0,\ldots,0)$, $n=1,\ldots,r\}$ enforce world-sheet 
supersymmetry; all are  mutually local. 

Simple currents preserving world-sheet and space-time supersymmetry are mutually local w.r.t. the 
Gepner model currents, see~\cite{Schellekens:1989wx}. $\mathfrak{J}=(0|2\rho_1,\ldots,2\rho_r|0,\ldots,0)$ 
is always local w.r.t. $\{\, J_n\, \}$, while mutual locality  
w.r.t. $J_0$ requires $\sum_n \rho_n /k_n \in \mathbb{Z}$. 
Extending a Gepner model with {\it all} such simple currents (without discrete torsion)  gives the mirror Gepner model, 
whose right R-charge has opposite sign. This is the basis of the construction of mirror 
manifolds by Greene and Plesser~\cite{Greene:1990ud}. 

\section{Non-geometric CY orbifolds}
We consider here simple currents that are {\it not} mutually local w.r.t. the Gepner model currents.  
A generic current $\mathfrak{J}$ as above is non-local w.r.t. $J_0$, 
hence space-time supersymmetry is broken. Now comes the key step; there is a choice of discrete torsion, consistent with 
eq.~(\ref{eq:torsionconst}) for any $\{\rho_n \in \mathbb{Z}\}$, bringing down the $X$ matrix to a lower-triangular form, 
whose only non-zero entries are
\begin{equation}
\label{eq:Xtriangle}
X_{\mathfrak{J}\mathfrak{J}} = \sum_{n=1}^r \frac{\rho_n^2}{k_n} \ , \quad X_{\mathfrak{J}0} = \sum_{n=1}^r \frac{\rho_n}{k_n} \, .
\end{equation}
Grouping the free-fermion and minimal models characters as 
$\chi^\lambda_\mu = \theta_{\sigma_0,2}/\eta \times \prod_{n=1}^r \chi^{j_n \phantom{(s_n)} }_{m_n,\, s_n}$,  
the modular-invariant partition function of the $\mathfrak{J}$-extended Gepner model is 
(with $K=\text{lcm} (4,2k_1,\ldots,2k_r)$ and $N=\text{lcm}\, \left( \text{lcm}\,(\rho_1,k_1)/\rho_1,\ldots,  
\text{lcm}\, (\rho_r,k_r)/\rho_r\right)$):
\begin{widetext}
\begin{multline}
Z = \frac{1}{2^r}\frac{1}{\tau_2^2 |\eta|^4}  \sum_{\lambda, \mu} \sum_{b_0 \in \mathbb{Z}_K} 
(-1)^{b_0}\ \delta^{(1)} \left(\frac{Q_R-1}{2} \right) \sum_{B \in \mathbb{Z}_N} 
\delta^{(1)} \left( \sum_{n=1}^r \frac{\rho_n (m_n+b_0 + \rho_n B)}{k_n}\right) \, \times \\ \times \, 
\prod_{n=1}^{r}  \sum_{b^n \in \mathbb{Z}_{2}}\delta^{(1)} \left(\frac{s_0-s_n}{2} \right)
\chi^\lambda_\mu (q) \chi^\lambda_{\mu + \beta_0 b_0 + \beta_l b^l + \beta_{\mathfrak{J}} B}(\bar q)\, \, . 
\label{eq:asycy3part}
\end{multline}
\end{widetext}
Thanks to the discrete torsion the projection onto odd-integer R-charges is restored in the left-moving sector. However twisted 
sectors associated with the $\mathfrak{J}$-extension ($B \neq 0$) have fractional right R-charge, hence space-time 
supersymmetry from the right is generically broken while supersymmetry from the left is preserved. This construction 
provides a whole class of non-geometric quotients of CY sigma-models at Gepner points. 

\section{Fractional mirror symmetry}
A particular type of  $\mathfrak{J}$-extensions with discrete torsion is quite interesting. Extending 
a Gepner model partition function with $\mathfrak{J}_1 = (0|2,0,\ldots,0|0,\ldots,0)$ amounts, while taking into account the 
twisted sectors and discrete torsion, to replace in the original partition function  the anti-holomor\-phic 
character for the first minimal model  with the character of opposite $\mathbb{Z}_{2k_1}$ charge, namely
\begin{equation}
\chi^{j_1}_{m_1 + b_0,\, \, s_1+b_0+2b_1 }(\bar q) \ \xrightarrow{\mathfrak{J}_1\text{\tiny -ext.}} 
\ \chi^{j_1}_{-m_1  - b_0,\, s_1+b_0+2b_1}(\bar q)\, .
\end{equation} 
In the right NS sector it is equivalent to change the sign of the right R-charge associated with the first minimal model; 
in the R sector  as well if one changes the right-moving space-time chirality at the same time. 
As superconformal field theories the original model and the new one are isomorphic, hence the two theories are 
dual\footnote{The R-current {\small 
$\bar{J}_R = - \bar{J}_{R\, ,(1)} + \sum_{\imath=2}^r  \bar{J}_{R\, ,(\imath)}$} has an integer spectrum, hence  can be 
exponentiated to a spin field mutually local with physical states. Hence there is another realization of supersymmetry in the dual model.}. 
Starting from a type IIA CY compactification at a Gepner point, we obtain a type IIB theory on a 
Gepner model whose right-moving R-charge associated with the first minimal model has been reversed; with respect to the 
original right-moving diagonal R-current the spectrum of R-charges is not integer-valued hence the model 
is not associated with a Calabi--Yau. Put it differently the quotient does not preserve the holomorphic three-form. 
These two models are isomorphic CFTs hence describe the same physics. 

A minimal model is the IR fixed point of a LG model 
with superpotential $W=Z^{k}$~\cite{Witten:1993jg}. Its mirror, obtained by a $\mathbb{Z}_k$ quotient, 
is a LG model for a twisted chiral superfield $\tilde{Z}$ with a twisted superpotential 
$\tilde{W}=\tilde{Z}^k$. In the present context we are considering a similar quotient acting inside a LG 
orbifold, with a discrete torsion that  disentangles partly the two orbifolds --~the diagonal one ensuring R-charge 
integrality and the $\mathbb{Z}_{k_1}$ quotient giving the fractional mirror. We end up with a model containing 
both a twisted chiral superfield $\tilde{Z}_1$ and chiral superfields $Z_{2,\ldots,r}$, hence cannot be related 
to a CY GLSM. This {\it fractional mirror symmetry} can be applied stepwise until one obtains 
the mirror Gepner model in the usual sense.

The asymmetric $K3\times T^2$ Gepner models that we have constructed in~\cite{Israel:2013wwa} can be rephrased 
in light of this new understanding. One considers two $\mathfrak{J}$-extensions acting respectively in the first and second factors  
of a $K3$ Gepner model,  and as $\mathbb{Z}_{k_{1}}$ and $\mathbb{Z}_{k_{2}}$ shifts along 
the two-torus. These models are close relatives of T-folds~\cite{Dabholkar:2002sy}, given that the $K3$ fiber is twisted 
by non-geometric symmetries as one goes around one-cycles of the  base. They interpolate between the $K3$ 
sigma--model in the large torus limit, and a 'half-mirror' K3 in the opposite small volume limit.

\section{Discussion}

We have considered new quantum symmetries associated with two-dimensional superconformal field theories lying  
the moduli space of Calabi--Yau compactifications. Usual T-dualities are associated with continuous symmetries corresponding to globally 
defined Killing vectors. Compact Calabi--Yau sigma-models have no continuous symmetries (apart from the axial and vector $U(1)_R$)   
but new dualities exist when they have accidental discrete symmetries.  Hence, unlike mirror symmetry, 
the fractional mirror map is a symmetry only at specific loci where these symmetries are manifest.  

To illustrate this point let us consider the quintic, giving the Gepner 
model $k_1=\cdots=k_5=5$. Away from the Gepner point, we expect that for every hypersurface 
\begin{equation}
z_1^5 + \sum \alpha_{abc} z_2^{a} z_3^{b} z_4^{c} z_5^{5-a-b -c}=0
\end{equation}
a fractional mirror w.r.t. the chiral superfield $Z_1$ exists. 
In other words the complex structure deformations that preserve the 
$\mathbb{Z}_{5}$ symmetry $z_1 \to e^{2i\pi/5} z_1$ are compatible with this duality. K\"ahler deformations are not 
allowed, as can be seen explicitly at the Gepner point, excluding the existence of a large-volume limit. 
When these conditions are met the $\mathcal{N}=2$ superconformal algebra can be split into the algebra coming from the LG model 
$W=Z_1^{\, 5}$ and from the LG model for the other multiplets. This allows to dualize $Z_1$ into a twisted chiral multiplet, 
giving a more general 'hybrid' LG orbifold with superpotential $W=\sum \alpha_{abc} Z_2^{a} Z_3^{b} Z_4^{c} Z_5^{5-a-b -c}$ 
and twisted superpotential $\tilde W = \tilde{Z}_1^5$. One expects also that other accidental splittings of the superconformal algebra 
should give rise to different fractional mirror symmetries. 

One may argue that, after all, the quotient is 'almost' geometric as the discrete torsion only plays a role in the twisted sectors. 
This is not actually correct, as the tensor product of minimal models becomes a CY sigma-model only after the Gepner orbifold has been implemented. 
The discrete torsion has an effect in the twisted sectors of the $J_0$-extension, giving the compactification a 
non-geometric nature. In particular the quotient has a different action on twisted chirals, $i.e.$ on K\"ahler moduli, compared to the corresponding geometric orbifold.

A more geometrical characterization of the fractional mirrors, that are not expected to be in any Calabi--Yau moduli space, would be very helpful in understanding their properties. 
Following the approach of Hori and Vafa~\cite{Hori:2000kt}, one may start with a Calabi--Yau GLSM and dualize only part of the chiral superfields. 

Finally it would be interesting to find whether these symmetries are related to the {\it Mathieu moonshine}, which  
suggests that $K3$ compactifications have an underlying $M_{24}$ symmetry whose origin is not fully understood~\cite{Eguchi:2010ej}.

\section{Acknowledgments}
\begin{acknowledgments}
I thank Ilka Brunner, Atish Dabholkar, Stefan Groot Nibbelink, Nick Halmagyi, Jan Louis, Michela Petrini and Jan Troost for 
discussions. This work was conducted within the ILP LABEX (ANR-10-LABX-63) supported by French state funds managed by the
ANR (ANR-11-IDEX-0004-02) and by the project QHNS in the program ANR Blanc SIMI5 of Agence National de la
Recherche. 
\end{acknowledgments}

\bibliography{bibmirror}

\end{document}